\documentclass[aps,prb,article,twocolumn,preprintnumbers,amsmath,amssymb,superscriptaddress]{revtex4-2}

\date{\today}
\usepackage{graphicx}
\usepackage{xcolor} 
\usepackage[colorlinks,citecolor=blue,linkcolor=blue,hyperindex]{hyperref}
\usepackage{float}
\usepackage{hyperref}
\usepackage{comment}
\usepackage{amsmath}
\usepackage{braket}
\usepackage{lineno}

\newcommand{\qbf}      {{q}}

\newcommand{\Ub}{U_{\rm GS}}
\newcommand{\Rq}{R_\qbf}
\newcommand{\Rqd}{R_\qbf^\dagger}
 
\hypersetup{
    colorlinks=blue,
    linkcolor=blue,
    filecolor=magenta,      
    urlcolor=blue,
    pdftitle={Overleaf Example},
    pdfpagemode=FullScreen,
    }

\begin{abstract}
\textbf{Abstract}:
Interactions between electrons and phonons play a crucial role in quantum materials. Yet, there is no universal method that would simultaneously accurately account for strong electron-phonon interactions and electronic correlations. By combining methods of the variational quantum eigensolver and the variational non-Gaussian solver, we develop a hybrid quantum-classical algorithm suitable for this type of correlated systems. This hybrid method tackles systems with arbitrarily strong electron-phonon coupling without increasing the number of required qubits and quantum gates, as compared to purely electronic models. We benchmark our method by applying it to the paradigmatic Hubbard-Holstein model at half filling, and show that it correctly captures the competition between charge density wave and antiferromagnetic phases, quantitatively consistent with exact diagonalization.
\end{abstract}

\begin{document}

\title{A Hybrid Quantum-Classical Method for Electron-Phonon Systems}
\author{M. Michael Denner}
\affiliation{Department of Physics, University of Zurich, Winterthurerstrasse 190, 8057 Zurich, Switzerland}
\author{Alexander Miessen}
\affiliation{IBM Quantum, IBM Research -- Zurich, 8803 Rüschlikon, Switzerland}
\affiliation{Institute for Computational Science, University of Zurich, Winterthurerstrasse 190, 8057 Zurich, Switzerland}
\author{Haoran Yan}
\affiliation{Department of Physics and Astronomy, Clemson University, Clemson, SC 29634, USA}
\affiliation{Department of Chemistry, Emory University, Atlanta, GA 30322, USA}
\author{Ivano Tavernelli}
\affiliation{IBM Quantum, IBM Research -- Zurich, 8803 Rüschlikon, Switzerland}
\author{Titus Neupert}
\affiliation{Department of Physics, University of Zurich, Winterthurerstrasse 190, 8057 Zurich, Switzerland}
\author{Eugene Demler}
\affiliation{Institute for Theoretical Physics, ETH Z\"{u}rich, 8093 Zürich, Switzerland}
\author{Yao Wang}
\email[\href{mailto:yao.wang@emory.edu}{yao.wang@emory.edu}]{}
\affiliation{Department of Physics and Astronomy, Clemson University, Clemson, SC 29634, USA}
\affiliation{Department of Chemistry, Emory University, Atlanta, GA 30322, USA}
\maketitle

\section{Introduction}

Understanding strongly correlated many-body systems is vital to many areas of science and technology, such as the development and analysis of functional quantum materials\,\cite{keimer2015quantum}. Due to the entanglement induced by correlations, macroscopic properties of quantum materials are often unpredictable from reductive single-particle models. Theoretical analysis of these systems with strongly entangled degrees of freedom has, however, been hindered by the exponential growth of their Hilbert space sizes with the number of particles. Understanding macroscopic properties of materials requires the analysis of sufficiently large model systems, which cannot be done accurately with classical computers. Quantum computing technologies, including hybrid quantum-classical algorithms\,\cite{peruzzo2014variational,yung2014transistor, farhi2014quantum} constitute an intriguing new direction for studying quantum many-body systems and especially quantum materials. 

\begin{figure*}[t]
    \includegraphics[width=\linewidth]{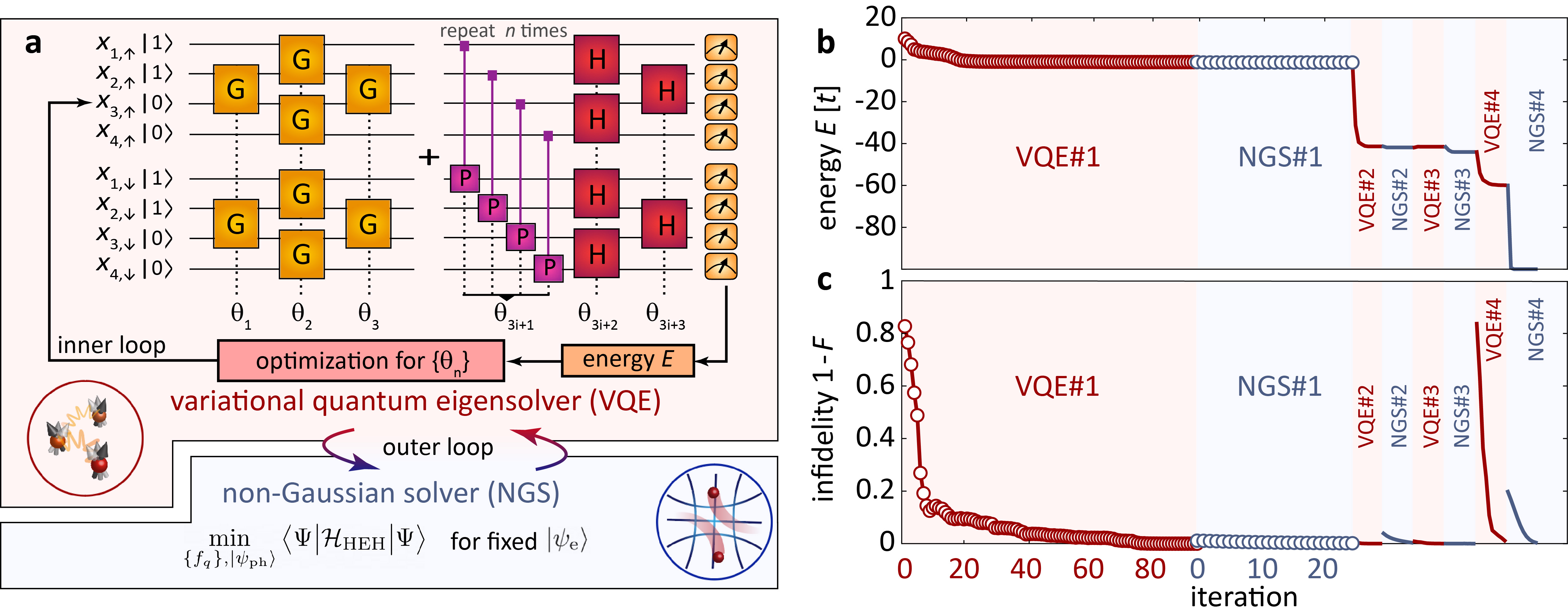}\vspace{-4mm}
    \caption{\label{fig:circuit_vqe} \textbf{Hybrid quantum algorithm.} \textbf{a} The hybrid quantum algorithm iterates between a variational quantum eigensolver (VQE) for the electronic and a non-Gaussian solver (NGS) for the phonon part of the many-body ground state. The quantum circuit structure for a 4-site example at half filling contains Givens rotations \textsf{G}, on-site gates \textsf{P}, and hopping gates \textsf{H}. The \textsf{P} and \textsf{H} layers are repeated $n$ times to express the ground state wavefunction. Within each layer, gates share the same variational parameters $\theta_i$, which are optimized on a classical computer inside each VQE iteration. \textbf{b} Convergence of the NGS-VQE algorithm, reflected by the total energy as a function of inner-loop (NGS or VQE) iteration steps for a 4-site Hubbard-Holstein model with $u = 10$, $\lambda = 10$, and $\omega = 1$. VQE steps were performed with quantum circuit statevector simulations and a circuit depth of $n=5$. Alternative outer-loop iterations are colored by red (for VQE) and blue (for NGS) and the data points are compressed after NGS \#1, for illustration purposes. \textbf{c} Convergence of the ground state infidelity $1-F$ during each iteration. The reference state chosen for each outer-loop iteration was obtained by exact diagonalization on classical computers.}
\end{figure*}

One of the promising representatives of hybrid algorithms is the variational quantum eigensolver (VQE)\,\cite{peruzzo2014variational, moll2018universal, endo2021hybrid,cerezo2021applications,bharti2022noisy}. This approach aims to accurately approximate ground states of quantum systems that can be naturally represented using qubits, such as spin and fermionic models. An example of such a protocol is shown in the upper panel of Fig.~\ref{fig:circuit_vqe}\textbf{a}: one uses a set of parameterized quantum gates to prepare a variational wavefunction and measures the expectation value of the Hamiltonian; then, one optimizes the parameters of these quantum gates using a classical computer. VQE has been implemented experimentally for small molecules\,\cite{kandala2017hardware, colless2018computation,takeshita2020increasing, motta2020determining,ollitrault2019quantum}, providing accurate solutions verifiable by exact methods. Hardware-efficient implementations of VQE have also been proposed for solid-state systems, including quantum magnets and Mott insulators\,\cite{uvarov2020variational,cade2020strategies,suchsland2022simulating, stanisic2022observing}. Successful applications of the VQE approach to describe multi-orbital molecules and intermediate-size solid-state models pave the way for extending this technique to broader classes of materials.

However, realistic materials usually contain more complex interactions than simplified electronic models, such as the Hubbard model which only features local Coulomb interaction. The interaction between mobile electrons and the ionic lattice in solids, so-called electron-phonon coupling (EPC), underlies electric and mechanical properties of materials. Notably, it has been suggested that the interplay between the electron-phonon interaction and the electronic Coulomb repulsion is crucial for many novel quantum phases, such as unconventional superconductivity in cuprates\,\cite{scalapino1986d,gros1987superconducting,shen2004missing, lanzara2001evidence, reznik2006electron, he2018rapid} and twisted bilayer graphene\,\cite{kang2019strong, hejazi2019multiple,guinea2018electrostatic,lian2019twisted,wu2018theory,peltonen2018mean}. Achieving predictive control of these quantum phases calls for developing reliable theoretical models describing materials with EPC\,\cite{fausti2011light,hu2014optically, boschini2018collapse}, which has motivated studies based on small clusters\,\cite{rosch2004electron,khatami2008effect, wang2018light} or perturbative couplings\,\cite{murakami2013ordered,kemper2015direct,sentef2016theory,babadi2017theory}. Quantum simulation of materials with strong EPCs, however, remains challenging due to the unbounded phonon Hilbert space~\cite{macridin2018electron,li2023efficient,macridin2018digital,pavosevic2021polaritonic}. The common spirit of quantum algorithms is to traverse quantum states encoded by the combination of available qubits. Therefore, even with a single electronic band and single phonon mode, the system has much higher computational complexity compared to electrons alone: inclusion of phonons in an $L$-site spinful system increases the Hilbert-space size from $4^L$ (for electrons) to $4^L(m+1)^L$ where $m$ is the (truncated) maximal local phonon occupation. For materials with non-negligible EPCs, the required $m\gg 1$ leads to an unreasonably large, and even unbounded Hilbert space. This issue prohibits not only classical simulations, but also an efficient encoding on a quantum machine.

To this end, we design a hybrid quantum algorithm which leverages the capability of VQE with quantum computers and the variational non-Gaussian description of non-perturbative polaronic dressing\,\cite{shi2018variational, wang2020zero,wang2021fluctuating}. We prove the validity of our approach using the one-dimensional Hubbard-Holstein model and its variants, which is summarized together with the specifics of the algorithm in the ``Methods'' section. We then show that our hybrid quantum algorithm is able to reliably capture the ground-state properties of the paradigmatic Hubbard-Holstein model in all regions of the phase diagram, when compared to non-Gaussian exact diagonalization (NGSED) results. Our algorithm does not require any additional qubit overhead stemming from unbounded phononic degrees of freedom and the truncation to a low phonon occupation\,\cite{macridin2018electron,macridin2018digital,pavosevic2021polaritonic}, as we implicitly sample the phonon Hilbert space. This makes it possible to analyze electron-phonon systems over a broad range of parameters, including both adiabatic and anti-adiabatic regimes of the phonon frequencies. We conclude our analysis by investigating the scaling of the algorithm's performance with respect to the system size, indicating the reduction of exponentially increasing complexity. Moreover, unlike prior studies on EPC systems implemented in trapped ions~\cite{casanova2011quantum,mezzacapo2012digital}, our approach is not limited to a particular hardware platform. This makes our algorithm a promising candidate for studying systems with strong EPC and electron-electron interactions beyond classically solvable problems.

\section{Results}
\subsection{Variational non-Gaussian VQE (NGS-VQE) method}
We consider a prototypical correlated system, where electronic correlations are described by the local Coulomb interaction -- the Hubbard model, while electron-phonon coupling follows the linear Fr\"ohlich-type density-displacement interaction.  While the latter is usually also simplified into local couplings -- the Holstein model -- recent experimental discoveries in cuprates have indicated the importance of nonlocal couplings\,\cite{chen2021anomalously, wang2021phonon, tang2022traces, wang2022spectral}. Including all these interactions, we obtain the Hubbard-extended-Holstein (HEH) model, whose Hamiltonian is
\begin{equation}
\begin{aligned}
\label{eq:HH_Hamiltonian}
    \mathcal{H}_{\rm HEH} = &-t \sum_{\langle i,j \rangle, \sigma} \left(c_{i,\sigma}^\dagger c_{j,\sigma}^{} + \mathrm{h.c.}\right) + U \sum_i n_{i,\uparrow}n_{i,\downarrow}\\
    &+ \sum_{i,j,\sigma}g_{ij} \left(a_i^{} + a_i^\dagger\right)n_{j,\sigma} + \omega_0 \sum_i a_i^\dagger a_i^{}. 
\end{aligned}
\end{equation}
Here, $c_{i,\sigma}^{}$ ($c_{i,\sigma}^{\dagger}$) annihilates (creates) an electron at site $i$ with spin $\sigma$, and $a_i^{}$ ($a_i^{\dagger}$) annihilates (creates) a phonon at site $i$; $n_{i,\sigma} = c_{i,\sigma}^\dagger c_{i,\sigma}^{}$ denotes the electron density operator for site $i$ and spin $\sigma$. Among the model parameters, $t$ sets the (nearest-neighbor $\langle i,j \rangle$) hopping integral, $U$ sets the on-site repulsive interaction, $\omega_0$ sets the Einstein phonon energy. $g_{ij}$ is the coupling strength between the phonon displacement at site $i$ and electron density at site $j$. While our method can tackle any distribution of EPCs, as discussed later, we restrict ourselves to local $g= g_{ii}$ and nearest-neighbor coupling $g'=g_{i,i\pm1}$ in one dimension. 

When the EPC is local, i.e., $g_{ij}=g\delta_{ij}$, the HEH model is reduced to the Hubbard-Holstein model. The physical properties of the Hubbard-Holstein model have been studied with various numerical methods, in one-dimensional (1D) systems\,\cite{hotta1997unconventional,fehske2002peierls,fehske2003nature, tezuka2007phase,fehske2008metallicity,ejima2010dmrg,clay2005intermediate,greitemann2015finite}, two-dimensional (2D) systems\,\cite{nowadnick2012competition,nowadnick2015renormalization,karakuzu2017superconductivity, ohgoe2017competition,hohenadler2019dominant}, and infinite dimensions\,\cite{georges1996dynamical, werner2007efficient, backes2023dynamical}. The phase diagram of the Hubbard-Holstein model is controlled by three dimensionless parameters, notably $u = U/t$ for electronic correlations, $\lambda = g^2/\omega_0 t$ for the effective EPC, and $\omega = \omega_0/t$ for phonon retardation effects. The presence of nonlocal EPCs has been studied recently, motivated by the observed attractive nearest-neighbor interactions in cuprate chains\,\cite{chen2021anomalously}. Due to this reason, numerical studies of the HEH model were primarily focused on 1D systems\,\cite{wang2021phonon, tang2022traces}. In this paper, we also restrict ourselves to periodic 1D systems, while the presented algorithm can be naturally extended to 2D.

To handle the strongly entangled electronic wavefunction and unbounded phonon Hilbert space simultaneously, we employ a variational non-Gaussian construction of the many-body wavefunction\,\cite{shi2018variational,shi2020variational}. A universal electron-phonon wavefunction can always be written in the form of
\begin{eqnarray}\label{eq:wvfuncansatzFull}
    |\Psi\rangle  = U_{\rm NGS}(\{f_q\}) |\psi_{\rm ph}\rangle \otimes|\psi_{\rm e}\rangle,
\end{eqnarray}
where the right-hand side is a direct product of electron and phonon states (denoted as $|\psi_{\rm e}\rangle$ and $|\psi_{\rm ph}\rangle$, respectively), with the variational non-Gaussian transformation $U_{\rm NGS}=e^{i\mathcal{S}}$ and the (Hermitian) operator $\mathcal{S}$ being a polynomial formed by $c$, $c^\dagger$, $a$, and $a^\dagger$ operators (with any sub-indices). The functional class of Eq.~\eqref{eq:wvfuncansatzFull} is a complete representation of the full electron-phonon Hilbert space. The variational solution based on this approach is guaranteed to be accurate, provided that $\mathcal{S}$ can assume arbitrary polynomials. The accuracy usually converges at a relatively low order in the exponent\,\cite{crawford2007introduction}. When determining the order of polynomials in $\mathcal{S}$, one should balance theoretical needs and computational feasibility: including high-order powers of electronic and phonon operators in $\mathcal{S}$ improve the accuracy and expand the applicability to complex models; however, these powers also lead to a large variational parameter space and a complex energy representation form. As benchmarked by exact diagonalization (ED) and determinant quantum Monte Carlo (DQMC) simulations of small clusters\,\cite{wang2020zero, wang2021fluctuating}, it is sufficient to truncate the $\mathcal{S}$ operator to the lowest-order terms for the Holstein-type coupling [see Eq.~\eqref{eq:S-operator} in the ``Methods'' section]. Denoting these lowest-order coefficients as $\{f_q\}$ ($q$ is the quantum number, naturally chosen as momentum for periodic systems), we have the variational non-Gaussian transformation $U_{\rm NGS}(\{f_q\})$ fully determined by these parameters.
 
Using this ansatz, we solve the HEH problem by minimizing the average energy
\begin{equation}\label{eq:groundE}
    E(\{f_q\}, |\psi_{\rm ph}\rangle, |\psi_{\rm e}\rangle) =\big\langle\Psi\big| \mathcal{H}_{\rm HEH} \big|\Psi\big\rangle
\end{equation}
self-consistently with respect to the unrestricted electronic state $|\psi_{\rm e}\rangle$ and the variational parameters in $U_{\rm NGS}(\{f_q\})$ and $|\psi_{\rm ph}\rangle$. Within each iteration, the variational non-Gaussian parameters $\{f_q\}$ and the phonon wavefunction $|\psi_{\rm ph}\rangle$ (here restricted to be a Gaussian state) are optimized using the imaginary-time equations of motion derived in Ref.~\onlinecite{wang2020zero} (see Fig.~\ref{fig:circuit_vqe}\textbf{a}). This is referred to as the non-Gaussian (NGS) solver, whose computational complexity scales polynomially with the system size $L$. On the other hand, the fully entangled electronic part $|\psi_{\rm e}\rangle$ of the wavefunction, represented by a tailored quantum circuit, can be obtained by regarding the $|\psi_{\rm ph}\rangle$ and $\{f_q\}$ as fixed and further minimizing the total energy. The latter step is equivalent to solving the electronic ground state of an effective Hubbard Hamiltonian $\mathcal{H}_{\rm eff} =  \langle\psi_{\rm ph}| U_{\rm NGS}^\dagger \mathcal{H}_{\rm HEH}U_{\rm NGS}|\psi_{\rm ph}\rangle$. Physically, this $\mathcal{H}_{\rm eff}$ describes the behavior of polarons, formed by phonon-dressed electrons. The phonon dressing gives rise to a heavier effective mass by modifying the hopping strength $\tilde{t}$ and mediates a long-ranged attraction $\tilde{V}_{ij}<0$ between polarons, in the form of Eq.~\eqref{eq:EH_Hamiltonian} in the ``Methods'' section. In the case of Holstein couplings, Gaussian states are an efficient representation of the phonon wavefunction $|\psi_{\rm ph}\rangle$\,\cite{shi2018variational,shi2020variational,wang2020zero}. Gaussian states allow to represent the effective nearest-neighbor hopping and phonon-mediated interactions in closed-form, as shown in Eqs.~\eqref{eq:effHopping} and~\eqref{eq:effInteraction} of the ``Methods'' section (see also Ref.~\onlinecite{wang2021phonon}).

Within each self-consistent iteration, the key complexity of solving the electron-phonon-coupled problem has thus been converted into solving a purely electronic Hamiltonian $\mathcal{H}_{\mathrm{eff}}$ with extended Hubbard interactions $\tilde{V}_{ij}$. This electronic problem can be efficiently embedded on a quantum hardware by using a suitable fermionic encoding. Here, we employ the Jordan-Wigner transformation, which maps each electron with given spin orientation to one qubit [see ``Methods'' section]. By applying a set of parameterized rotations to these qubits, we obtain a quantum circuit representing a variational electronic wavefunction $|\psi_{\rm e} (\{\theta_i\})\rangle$. The self-consistent quantum-classical iterations in VQE then optimize the variational gate parameters $\{\theta_i\}$ to minimize the energy $\langle \psi_{\rm e} (\{\theta_i\}) | \mathcal{H}_{\mathrm{eff}}|\psi_{\rm e} (\{\theta_i\})\rangle$. The solution of VQE approximates $|\psi_{\rm e} (\{\theta_i\})\rangle$ in the variational ground state of Eq.~\eqref{eq:wvfuncansatzFull}. Unless explicitly specified otherwise, we conduct the VQE step of the NGS-VQE iterations with exact statevector simulations using Qiskit~\cite{Qiskit}.

The standard Hubbard model has been successfully studied with various quantum circuits. Here, we employ the Hamiltonian variational ansatz\,\cite{wecker2015progress}, which can be naturally extended to nonlocal interactions: inspired by the quantum adiabatic theorem, it starts from an efficient encoding of the noninteracting lattice model with translational symmetry\,\cite{ortiz2001quantum,wecker2015solving,verstraete2009quantum,jiang2018quantum}, and then evolves the state using alternating kinetic energy and interacting terms\,\cite{cade2020strategies,stanisic2022observing} [see ``Methods'' section]. The specific ansatz for the quantum circuit is presented in the upper panel of Fig.~\ref{fig:circuit_vqe}\textbf{a}. Since the ground state of a finite-size periodic system preserves translational symmetry, we assume the quantum gates in the same layer to share the same parameters (denoted as $\theta_n$ in Fig.~\ref{fig:circuit_vqe}\textbf{a}). The expressibility of the quantum circuit ansatz is controlled by the repetition number $n$ of the evolution block ($\textsf{P}$ and $\textsf{H}$ gates in Fig.~\ref{fig:circuit_vqe}\textbf{a}). Depending on the hardware specific transpilation, the number of CNOT gates for a given $n$ follows as $6 + 5\times n$. This quantum circuit represents an efficient encoding if $n$ does not scale exponentially with the system size $L$. We investigate the scaling with the system size in the ``Scaling in circuit depth and system size'' subsection, and show that the chosen ansatz leads to a quantitatively accurate performance of the hybrid quantum algorithm. 

\begin{figure*}[!ht]
    \includegraphics[width=\linewidth]{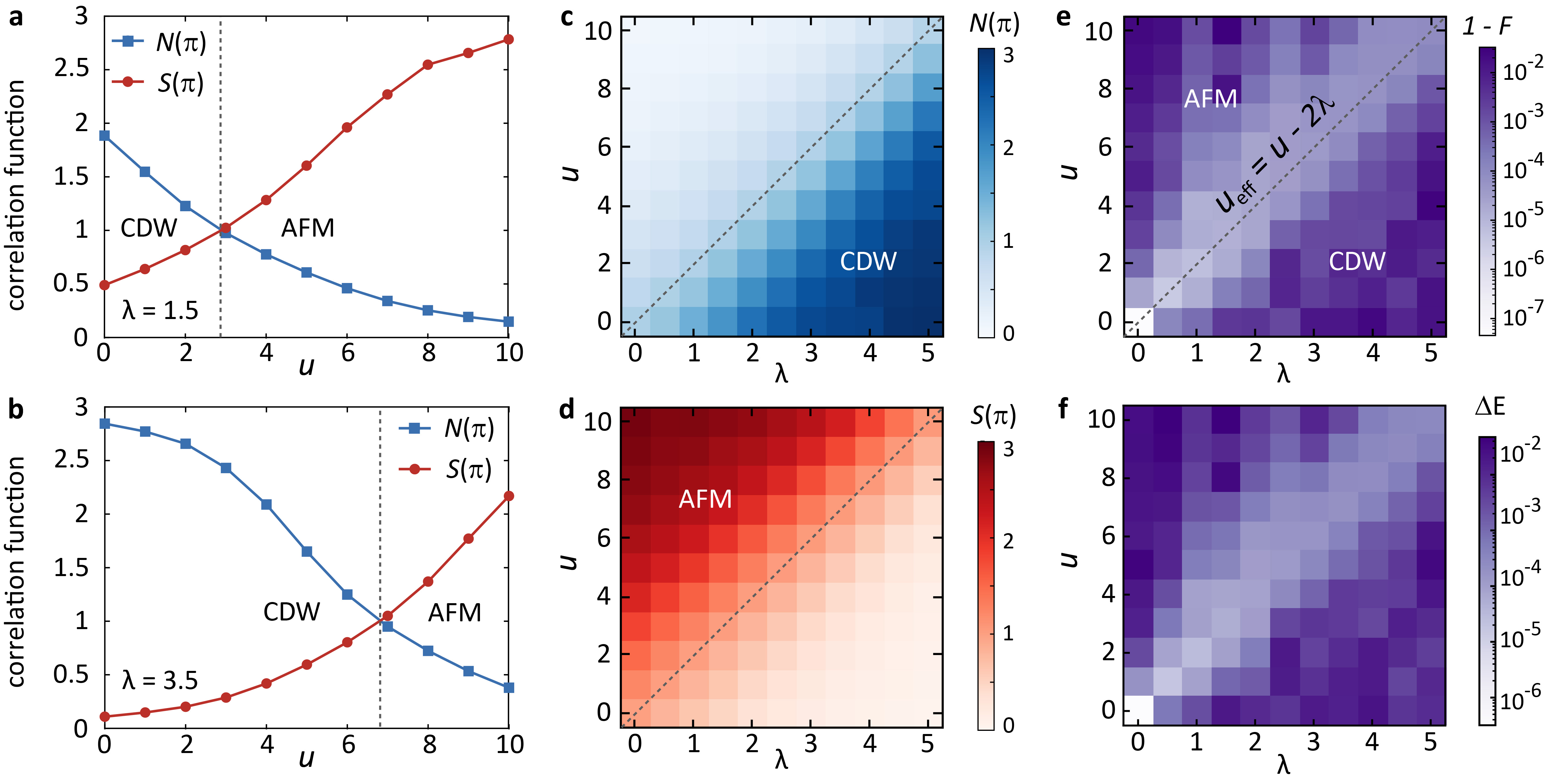}\vspace{-4mm}
    \caption{\label{fig:phase_diagram} \textbf{Phase diagram of the one-dimensional Hubbard-Holstein model.} \textbf{a}, \textbf{b} Charge $N(\pi)$ and spin $S(\pi)$ structure factors for the ground state of a 6-site Hubbard-Holstein model, simulated by the non-Gaussian solver variational quantum eigensolver (NGS-VQE) algorithm as a function of $u$ for fixed \textbf{a} $\lambda = 1.5$ and \textbf{b} $\lambda = 3.5$. The charge-density-wave (CDW) and antiferromagnetic (AFM) regimes are marked. The phonon frequency is set as $\omega = 10$ and the quantum circuit depth is $n = 9$. \textbf{c}, \textbf{d} Distribution of the static \textbf{c} charge and \textbf{d} spin structure factors in the $u-\lambda$ parameter plane, for the same conditions as \textbf{a} and \textbf{b}. The dashed line indicates the anti-adiabatic phase boundary $u = 2 \lambda$ between AFM and CDW. \textbf{e} Ground state infidelity and \textbf{f} (absolute-value) energy error with respect to ED results for the phase diagram in \textbf{c} and \textbf{d}.}
\end{figure*}

Figure \ref{fig:circuit_vqe}\textbf{b} shows an example of the NGS-VQE simulation for a 4-site Hubbard-Holstein model with $u=10$ and $\lambda=10$. Due to the variational nature of each step in the self-consistent NGS-VQE iteration, the total energy $E$ always decreases. Thus, this NGS-VQE iteration is ensured to converge to a local energy minimum within the variational space set by Eq.~\eqref{eq:wvfuncansatzFull}, which provides a good approximation for the ground state. It is worth noting that the energy evolution is not always smooth, where a sudden drop indicates a dramatic change in the variational wavefunction from one phase to another. In the example presented in Fig.~\ref{fig:circuit_vqe}\textbf{b}, the initial state is prepared by setting the phonon wavefunction to vacuum. Consequently, the first outer-loop iteration with VQE (VQE\#1 in the panel) starts with a pure Hubbard model, followed by the adjustment of phonon states and NGS parameters (NGS\#1 in the panel). After the entire first outer-loop iteration (VQE\#1 and NGS\#1), the electronic state $|\psi_{\rm e}\rangle$ lies in an antiferromagnetic (AFM) state as a solution for the pure Hubbard model, while the phonon state $|\psi_{\rm ph}\rangle$ induces a large attractive potential in the form of Eq.~\eqref{eq:effInteraction} in the ``Methods'' section. This phonon-mediated interaction tends to stabilize a charge density wave (CDW), which contradicts the AFM state (see discussion in ``Charge and spin phases in the Hubbard-Holstein model'' for details). As a result, the electronic state rapidly evolves once the second self-consistent iteration (VQE\#2) starts. 

In addition to the energy evolution, we parameterize the wavefunction error using the infidelity, defined as $1-F$ with the fidelity $F = \left|\langle \Psi_{\rm VQE} | \Psi_{\rm ED}\rangle\right|^2$. In this context, the reference ground state $| \Psi_{\rm ED}\rangle$ is chosen as the optimal solution for each VQE and NGS step of the outer-loop iteration. Figure \ref{fig:circuit_vqe}\textbf{c} shows the convergence of the wavefunction for the same system as Fig.~\ref{fig:circuit_vqe}\textbf{b}. By comparing these two panels, one can observe that a slow energy evolution may come with a relatively rapid change of wavefunction parameters, indicative of a barren plateau\,\cite{uvarov2020variational}. Therefore, the infidelity may experience significant changes in later (outer-loop) iterations when the energy is close to convergence. We emphasize that the infidelity is posterior and cannot be used as the target function of the iteration.  

In contrast to solving a Hubbard model, the NGS-VQE method involves a self-consistent outer loop between electrons and phonons. Thus, the combined NGS-VQE efficiency is essential for optimal results. To mitigate optimization issues of the variational quantum circuit, for instance the barren plateau or a multitude of local minima, we employ a three step optimization. First of all, since all gates within a single ansatz layer share the same variational parameter $\theta_i$, we can reuse parameters across circuits for different system sizes $L$. We therefore pre-run the VQE with smaller system sizes to initialize the circuit of the target system with these converged variational parameters. Moreover, the ground state evolves adiabatically for small changes in the model parameters ($u$, $\lambda$, and $\omega$) within the same phase. This is why we further recycle converged parameters if results for similar model parameters exist. Finally, we adaptively adjust the VQE convergence criterion during the outer-loop NGS-VQE iterations by increasing the number of quantum measurements. Since the initial phonon state and variational non-Gaussian parameters are far from saddle points, we start using just a small number of measurements to give a low-accuracy estimation of the electronic ground state; with the progress of NGS-VQE iterations, we gradually raise the convergence criterion for the electronic state. All these strategies help to improve the overall performance and reduce the runtime of the hybrid NGS-VQE algorithm, especially relevant in hardware implementations (see \hyperref[Supplementary_Note_1]{Supplementary Note 1}).

\subsection{Simulating the correlated electron-phonon systems}

\begin{figure*}[t!]
    \includegraphics[width=\linewidth]{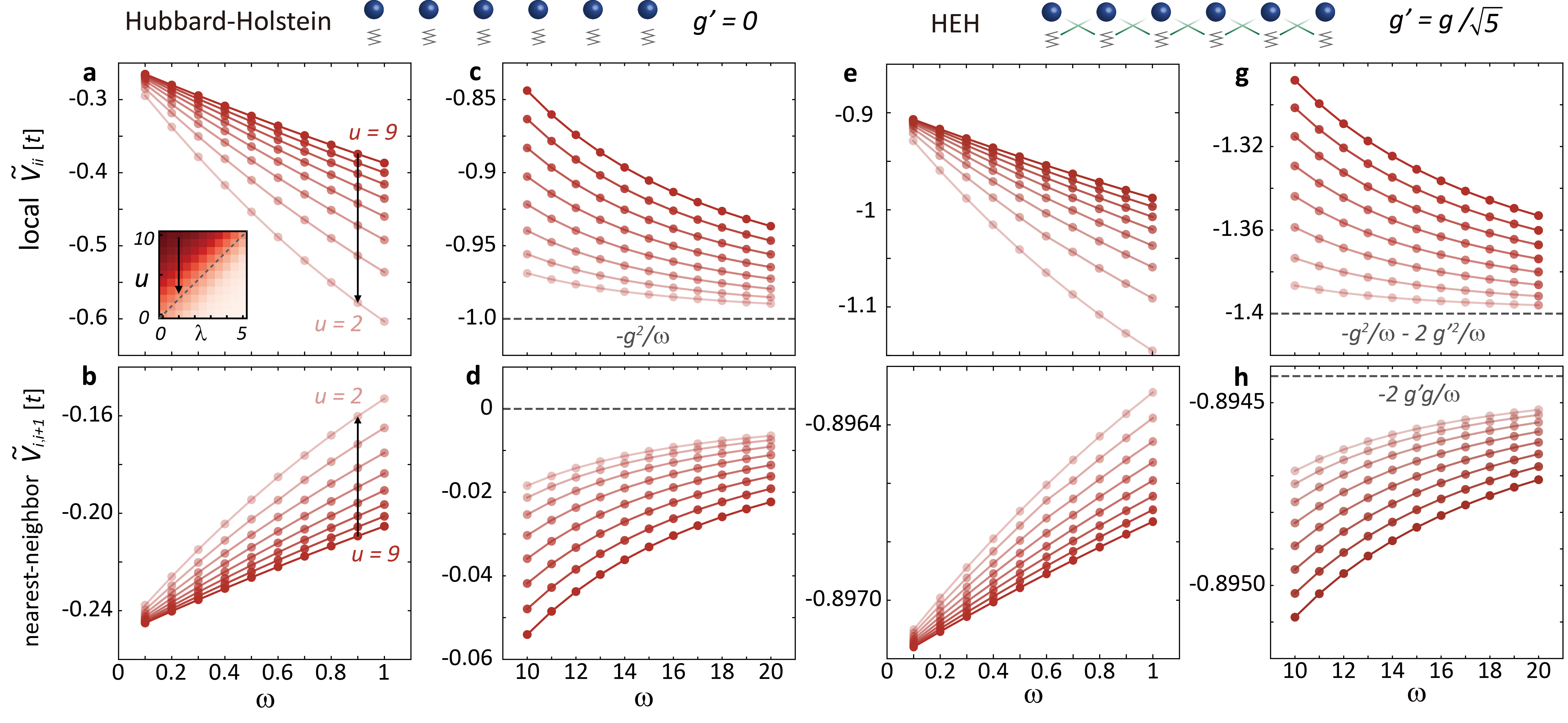}\vspace{-3mm}
    \caption{\label{fig:V_vs_g} \textbf{Phonon-mediated interactions for the ground state of the Hubbard-Holstein and Hubbard-extended-Holstein (HEH) model.} \textbf{a},\textbf{b} Local ($r = 0$) and nearest-neighbor ($r = 1$) interaction strengths $\tilde{V}$ for various Hubbard $u$ values within the antiferromagnetic (AFM) phase ($g' = 0, \lambda = 1$) and small phonon frequencies (close to the adiabatic limit). 
    The inset in \textbf{a} highlights the considered path in the phase diagram (black arrow). \textbf{c}, \textbf{d} Local ($r = 0$) and nearest-neighbor ($r = 1$) interaction strengths $\tilde{V}$ for the same interaction parameters as \textbf{a} and \textbf{b}, but with large phonon frequencies. The dashed lines suggest the asymptotic values in the anti-adiabatic limit. 
    \textbf{e}-\textbf{h} Local ($r = 0$) and nearest-neighbor ($r = 1$) interaction strengths $\tilde{V}$ for the same conditions as (\textbf{a}-\textbf{d}), but for the extended Hubbard model with $g' = g/\sqrt{5}$. 
    Similarly, the dashed lines in \textbf{g} and \textbf{h} indicate the asymptotic values in the anti-adiabatic limit.
    }
\end{figure*}

\subsubsection{Charge and spin phases in the Hubbard-Holstein model}
The Hubbard-Holstein model and its extension set the stage to study the interplay of electronic correlations and EPCs in quantum materials. At half-filling and in 1D, this model results in a rich phase diagram, hosting an AFM, CDW, and a narrow intermediate phase\,\cite{hotta1997unconventional,fehske2002peierls,fehske2003nature, tezuka2007phase,fehske2008metallicity,ejima2010dmrg,clay2005intermediate,greitemann2015finite}. To demonstrate the accuracy and efficiency of the NGS-VQE algorithm, we first restrict ourselves to the pristine Hubbard-Holstein model with $g'=0$ and simulate the spin and charge structure factors of the ground state for different model parameters. The (static) spin structure factor is defined as
\begin{equation}
    S(q) = \sum_{ij} \langle  (n_{i\uparrow}- n_{i\downarrow}) (n_{j\uparrow}- n_{j\downarrow})\rangle e^{-i q\cdot (r_i-r_j)}/L,
\end{equation}
while the (static) charge structure factor is defined as
\begin{equation}
    N(q) = \sum_{ij} \langle  (n_{i\uparrow}+ n_{i\downarrow}) (n_{j\uparrow}+ n_{j\downarrow})\rangle e^{-i q\cdot (r_i-r_j)}/L.
\end{equation}
Using the half-filled system as the benchmark platform in this paper, we focus on the nesting momentum $q=\pi$ for both structure factors. In the regime where electronic interactions dominate ($u \gg \lambda$), the spin structure factor $S(\pi)$ prevails over the charge structure factor, reflecting an AFM state in a finite cluster (see Fig.~\ref{fig:phase_diagram}\textbf{a}, \textbf{b} for $\omega = 10$). With the increase of $u-2\lambda$, $N(\pi)$ gradually vanishes as charge degrees of freedom are frozen with a substantial energy penalty for double occupations. In the other limit where EPCs dominate ($\lambda \gg u$), the charge structure factor $N(\pi)$ prevails over the spin structure factor $S(\pi)$. This reflects the onset of a CDW state, although a more rigorous identification requires either scaling to larger system sizes or excited-state analysis. Physically, the CDW is stabilized by the energy gain through a lattice distortion, forming an alternating pattern of holons and doublons. We summarize the dependence of both spin and charge structure factors on the two interaction parameters in Figs.~\ref{fig:phase_diagram}\textbf{c} and \textbf{d}. The trends of these two observables reflect the two dominant phases qualitatively consistent with physical intuition. Due to the underlying finite-size system, the two phases are separated by a crossover instead of a sharp phase boundary. Recent studies have shown the presence of an intermediate Luther-Emery liquid phase for $u \approx 2 \lambda$, whose width is controlled by the phonon frequency\,\cite{greitemann2015finite}. The discussion of this phase requires finite-size scaling and is beyond the scope of this paper.

To determine the quantitative accuracy of our hybrid quantum algorithm, we compare the converged ground state for each set of model parameters against the NGSED solutions. The accuracy of the latter has been benchmarked by ED and DQMC\,\cite{wang2020zero, costa2020phase, wang2021fluctuating}. As shown in Fig.~\ref{fig:phase_diagram}\textbf{e}, the infidelity map suggests that even the largest error is in the single-digit percentage range, at most $0.03$. These errors do not change significantly with increasing system size, as outlined in the ``Scaling in circuit depth and system size'' subsection. Interestingly, the most accurate solutions (with infidelity of order $10^{-5}$) are obtained near the boundary of the CDW and AFM phases, i.e., along the diagonal $u \approx 2 \lambda$. In this regime, the finite-system solution is more metallic, due to the delicate balance between the electronic repulsion and phonon-mediated attraction. Therefore, the true ground state of systems near the phase boundary can be efficiently captured with a Slater determinant prepared by Givens rotations\,\cite{ortiz2001quantum,wecker2015solving,verstraete2009quantum,jiang2018quantum}. In contrast, the infidelity increases when the system evolves into CDW or AFM states, although the NGS-VQE algorithm yields quantitatively accurate results throughout the phase diagram. This observation is contrasting the intuition that the AFM or CDW states are more classical. Instead, these states are cat states in these small and low-dimensional systems. An accurate representation of these highly-entangled states with long-range correlations, therefore, requires deeper quantum circuits and, accordingly, more gates. The dependence on circuit depth will be discussed in the ``Scaling in circuit depth and system size'' subsection. This sensitivity of the simulation accuracy to the model parameters is also reflected by the error of the ground-state energy, as shown in Fig.~\ref{fig:phase_diagram}\textbf{f}.

Up to now, the benchmark has been conducted with relatively large phonon frequencies $\omega=10$, where the competition between CDW and AFM states is primarily controlled by the effective local interaction $u_{\rm eff} = u - 2\lambda$ after integrating phonon fields. However, phonon frequencies in typical correlated materials are usually comparable to the electronic bandwidth, if not even reaching the adiabatic limit ($\omega \rightarrow 0$). The dependence of charge and structure factors on different phonon frequencies is discussed in Supplementary Note 2. In the thermodynamic limit, smaller phonon frequencies usually lead to a steeper crossover between the two phases\,\cite{fehske2002peierls,fehske2003nature, tezuka2007phase,fehske2008metallicity}, with both $S(\pi)$ and $N(\pi)$ dropping more rapidly when approaching the phase boundary. Note, however, that this intermediate phase cannot be resolved in a small cluster.

\subsubsection{Phonon-mediated interactions in the Hubbard-extended-Holstein model}

The wavefunction ansatz in Eq.~\eqref{eq:wvfuncansatzFull} allows to extract the effective model $\mathcal{H}_{\rm eff}$ in the polaronic basis while solving for the ground state. This approach has been used to quantify the recently discovered nearest-neighbor electronic attraction $\tilde{V}$ in cuprate chains\,\cite{wang2021phonon}. Here, we evaluate $\tilde{V}$ for different systems using the NGS-VQE algorithm both to provide intuition about phonon-mediated interactions in different limits and to benchmark the validity of the method under different conditions.

We first examine the Hubbard-Holstein model without $g'$. Figures \ref{fig:V_vs_g}\textbf{a},\textbf{c} and \textbf{b},\textbf{d} shows the simulated local and nearest-neighbor attractive interaction in the polaronic basis. Since this interaction was discovered in cuprates with strong electron correlations, we restrict ourselves to the AFM regime ($u\gg\lambda$) and set $\lambda=1$. In the anti-adiabatic limit, the local interaction $\tilde{V}_{ii}$ asymptotically approaches $-\lambda = -1$ and all nonlocal interactions $\tilde{V}_{i\neq j}$ vanish (see Fig.~\ref{fig:V_vs_g}\textbf{c}, \textbf{d}). This is consistent with the integration of phonons in field theory. With the decrease of $\omega$ and proximity to $u$, Coulomb interactions start to influence the distribution of $\tilde{V}_{ij}$. Such an influence is more obvious for large $u$, where the electronic correlations are strong and renormalize the phonon self-energy. This effect can be captured by the wavefunction ansatz of Eq.~\eqref{eq:phononwvfuncansatz}. The electron-dressing effect for the phonon self-energy is described by $\Omega_{\textit{eff}}$ in ref.\cite{wang2020zero}. Simultaneously, the retardation effect of finite-frequency phonons mediates the effective interaction at finite distance. Therefore, the effective nearest-neighbor attraction $\tilde{V}_{i,i+1}$ increases for lower phonon frequencies (see Fig.~\ref{fig:V_vs_g}\textbf{b}, \textbf{d}).

\begin{figure*}[ht!]
    \includegraphics[width=\textwidth]{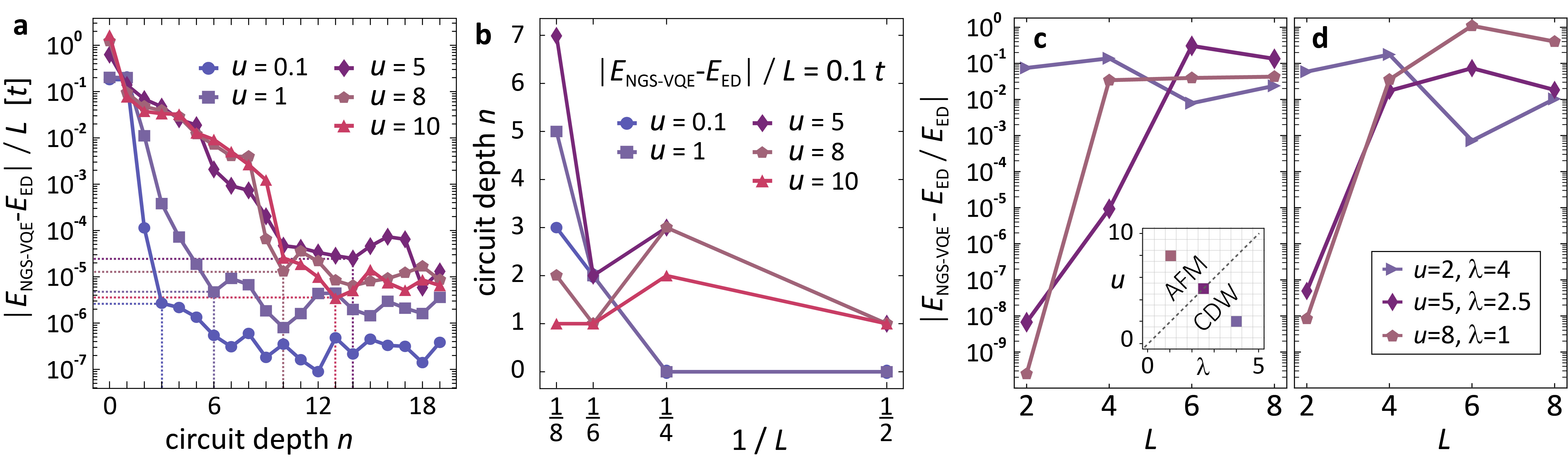}\vspace{-3mm}
    \caption{\label{fig:ansatz_errors} \textbf{Scaling behavior of the hybrid quantum algorithm.} \textbf{a} Simulation error for the ground-state energy $E_{\mathrm{NGS-VQE}}$ of the employed quantum circuit ansatz as a function of circuit depth $n$ and various on-site interactions $u$, compared to ED ($\lambda = 0$). This circuit depth $n$ follows the definition in Fig.~\ref{fig:circuit_vqe}\textbf{a} for a 6-site Hubbard model. \textbf{b} Scaling of circuit depth $n$ with system size $L$ for a fixed accuracy of $|E_{\mathrm{NGS-VQE}}-E_{\mathrm{ED}}|/L = 0.1t$ and various on-site interactions $u$ ($\lambda = 0$). The required circuit depth changes only slightly with the system size for small and large values of $u$. \textbf{c} Relative error of the converged ground-state energy for three distinct parameter sets for the Hubbard-Holstein model as a function of system size $L$ [the inset highlights the charge density wave (CDW) and antiferromagnetic (AFM) phase]. The phonon frequency is $\omega = 10$ and the variational quantum eigensolver (VQE) was performed with the circuit depth obtained in panel \textbf{b}. The error remains approximately constant when increasing $L$ from 4 to 8. \textbf{d} Same as panel \textbf{c}, but for a small phonon frequency $\omega = 1$. }
\end{figure*}

With further decreasing phonon frequencies, the impact of phonons becomes a mean-field-like deformation potential instead of a virtual process. Such a deformation potential contributes as an overall chemical potential instead of media of a two-particle interaction. Therefore, the simulated $\tilde{V}_{ij}$ becomes extremely nonlocal and evolves into an all-to-all interaction, which is equivalent to a chemical potential shift in a canonical ensemble with fixed particle number. In the $\omega\rightarrow0$ limit, phonon-mediated interactions are insensitive to the Hubbard $u$, because the mean-field phonon distortion is determined only by the average local electron density, which is approximately fixed in the AFM phase (see Fig.~\ref{fig:V_vs_g}\textbf{a}, \textbf{b}). This insensitivity is similar to the anti-adiabatic limit but has a different origin.

We now move on to discuss the impact of nearest-neighbor EPC $g'$ on the complexity and accuracy of the hybrid quantum algorithm. We fix the ratio $g'/g=1/\sqrt{5}$ to reflect the geometric relation of apical oxygens in a 1D transition-metal oxide\,\cite{wang2021phonon}. With electrons coupled to their nearest-neighbor phonons directly, the mediation of $\tilde{V}_{i,i+1}$ can be generated without retardation effect. Therefore, the asymptotic $\tilde{V}_{i,i+1}$ in the anti-adiabatic limit is no longer zero but acquires a finite value (see Fig.~\ref{fig:V_vs_g}\textbf{h}). This asymptotic interaction strength can be analytically calculated as $-2gg'/\omega$, which is more evident than the additional interactions caused by retardation effects (the latter is three orders of magnitude smaller than the former). At the same time, the phonon-mediated local interaction $\tilde{V}_{ii}$ is further strengthened by this nonlocal $g'$ coupling. The anti-adiabatic value of $\tilde{V}_{ii}$ approaches $-g^2/\omega - 2g'^2/\omega$ (see Fig.~\ref{fig:V_vs_g}\textbf{g}). Due to the geometric distance controlling $g'/g$, the strength of $\tilde{V}_{ii}$ is still comparable to that of the Hubbard-Holstein model ($g'=0$). In the adiabatic limit, the effective interactions are similar to those obtained from the Hubbard-Holstein model, asymptotically approaching an extremely delocalized $\tilde{V}_{ij}$. Compared to the $g'=0$ case (Figs.~\ref{fig:V_vs_g}\textbf{a} and \textbf{b}), the only difference is the asymptotic value when $\omega$ approaches zero. Using the fact that, in the long-wavelength limit ($q=0$), $\tilde{V}$ is proportional to $g_q^2/\omega$ and $g_q = g + 2g'\cos q$, we can estimate the ratio between these two asymptotic $\tilde{V}_{ij}$ values (for $g'=g/\sqrt{5}$ and $g'=0$) to be $(1+2/\sqrt{5})^2\approx 3.59$. This ratio agrees with the simulated results in Figs.~\ref{fig:V_vs_g}\textbf{a}, \textbf{b}, \textbf{e}, and \textbf{f}.

\subsection{Scaling in circuit depth and system size}\label{sec:scaling}

The results presented in Fig.~\ref{fig:V_vs_g} demonstrate that our hybrid quantum algorithm is able to produce quantitatively accurate results across the full phonon spectrum (see \hyperref[Supplementary_Note_2]{Supplementary Note 2} for a comparison to NGSED results). To further analyze the accuracy of our algorithm, we investigate the influence of different system sizes $L$. The depth of the quantum circuit controls the expressibility of the variational state, potentially allowing for a more accurate approximation of the electronic ground state by increasing $n$. However, an efficient encoding on quantum computing platforms is only possible if the depth of the employed quantum circuit does not scale exponentially with the system size $L$. As shown in Fig.~\ref{fig:ansatz_errors}\textbf{a}, the ground states for small-$u$ systems can be efficiently expressed by a Slater determinant. Thus, only a few layers are needed to reach ground state energy errors below $1\times10^{-6}$ when compared to ED. Larger $u$, however, requires deeper circuits, reaching a plateau of errors of the order $10^{-4}$ to $10^{-5}$ in the ground state energy. In order to investigate the scaling of necessary $n$ with the system size $L$, we consider a fixed error in the electron ground state energy of $|E_{\mathrm{VQE}}-E_{\mathrm{ED}}|/L = 0.1t$. The circuit depth required to achieve this performance as a function of $L$ is shown in Fig.~\ref{fig:ansatz_errors}\textbf{b}, highlighting a moderate increase in depth for small and large $u$. Intermediate values for $u$, however, require significantly deeper circuits, as quantum fluctuations are larger around the crossover between metallic and AFM phase.

So far, we have considered the influence of finite-depth quantum circuits on the electronic part of the many-body ground state. The hybrid NGS-VQE algorithm combines the electron solver with a variational NGS solver for the phonon and non-Gaussian components. That being said, errors in the VQE solutions do not necessarily accumulate, but can actually be mitigated by the phonon solver. Considering again a fixed error in the electron ground-state energy of $|E_{\mathrm{NGS-VQE}}-E_{\mathrm{ED}}|/L = 0.1t$ and corresponding circuit depth $n$, we investigate the accuracy of the hybrid quantum algorithm in different regions of the phase diagram. Specifically, we consider the relative error in the ground-state energy of the converged extended-Hubbard Hamiltonian [see Eq.~\eqref{eq:EH_Hamiltonian}], containing the phonon dressing of kinetic hopping and long-range interactions. Figures~\ref{fig:ansatz_errors}\textbf{c}, \textbf{d} indicate that the NGS-VQE simulation errors are usually at least an order of magnitude smaller than those of the electronic solvers. The largest errors are obtained for small phonon frequencies (Fig.~\ref{fig:ansatz_errors}\textbf{d}, $\omega = 1$), where the absence of quantum fluctuations hinders the phonon solver to escape local minima during the self-consistent iteration\,\cite{wang2020zero}. Warm-up iterations with larger phonon frequencies can help to alleviate this issue~\cite{wang2020zero}. Moreover, the relative errors do not increase for systems larger than $L = 4$, indicating quantitatively accurate results across different system sizes and phases. The only exception appears in the CDW phase at small phonon frequencies such as $\omega = 1$, where the relative error oscillates with $L$, likely due to the degeneracy of ground states. The ability to mitigate errors of the quantum solver also provides a promising path to experimental realizations. Hardware implementations, irrespective of the specific platform, suffer from decoherence\,\cite{deLeon2021materials}, making noise resilient algorithms of key importance. Our hybrid quantum algorithm is able to improve VQE results over a wide range of phonon frequencies, phase regions, and noise levels, suggesting efficient hardware realizations (see \hyperref[Supplementary_Note_3]{Supplementary Note 3}).

\section{Conclusions}
 
Our NGS-VQE method provides a general framework for performing accurate and efficient quantum simulations of electron-phonon systems with arbitrary interaction strengths. Using this method, we have studied the Hubbard-Holstein and HEH models as examples, reproduced the CDW-AFM crossover with high precision, and extracted the phonon-mediated interactions in a wide range of phonon frequencies. While we focused on paradigmatic (and experimentally relevant) cases, this method can be generally applied to any model with electronic Coulomb correlations and Fr\"ohlich-type electron-phonon couplings. The commutation between the NGS transformation [$e^{i\mathcal{S}}$ with $\mathcal{S}$ defined in the form of Eq.~\eqref{eq:S-operator} of the ``Methods''] guarantees a closed-form effective Hamiltonian similar to Eq.~\eqref{eq:EH_Hamiltonian}. This hybrid algorithm can be extended to other types of electron-boson interactions (like the Su-Schrieffer-Heeger phonon and cavity QED) through the generalization of the non-Gaussian transformation $U_{\rm NGS}$ and its optimization strategy. Anharmonic potentials can be tackled at the price of replacing $|\psi_{\rm ph}\rangle$ by more complicated many-body wavefunctions similar to the electronic ones. Both generalizations are accompanied by the increase of computational complexity and should be designed based on the requirements of specific models. Moreover, as demonstrated in Ref.\,\cite{wang2021fluctuating}, this framework can be extended to non-equilibrium dynamics, which requires a reliable quantum solver for the long-time propagation of $|\psi_{\rm e}\rangle$. The Fourier transform of non-equilibrium dynamics with two- or multi-time correlation functions further paves the way to excited-state spectra\,\cite{miessen2023among,shi2019ultrafast, white2004real}.

Our work shows that the phonon solver can mitigate potential errors of the quantum hardware, facilitating a future experimental implementation. The success of an experimental realization, however, relies on an efficient implementation of the required quantum circuits, respecting the connectivity of a given device. This is especially the case for nonlocal gates, like the $\textsf{P}$ gates of our ansatz representing electron-electron interactions.
Trapped ion and cold atom based platforms offer high qubit-connectivities~\cite{linke2017experimental}, reducing potential swap-overheads when implementing entangling gate-layers such as the $\textsf{P}$ and $\textsf{H}$ gates in our ansatz. Superconducting systems, on the other hand, have a more limited qubit-connectivity but operate at much faster rates, making them favorable for two-level NGS-VQE iterative schemes. Therefore, practical implementations of the algorithm  proposed in this study call for developing higher-connectivity superconducting hardware or error mitigation schemes to compensate for potential swap-overheads.

\section*{Methods}
\subsection{NGS-VQE method and effective Hamiltonian}
As mentioned in the main text, the variational electron-phonon wavefunction in the NGS-VQE method is given as 
\begin{eqnarray}\label{eq:wvfuncansatzFullm}
    \big|\Psi\big\rangle  = U_{\rm NGS}(\{f_q\}) |\psi_{\rm ph}\rangle \otimes|\psi_{\rm e}\rangle,
\end{eqnarray}
with the non-Gaussian transformation $U_{\rm NGS}=e^{i\mathcal{S}}$. As benchmarked by ED and DQMC on small clusters\,\cite{wang2020zero, wang2021fluctuating}, it is sufficient to truncate the $\mathcal{S}$ operator to the lowest-order terms
\begin{equation}
    \mathcal{S}(\{f_q\}) = -\frac{1}{\sqrt{L}}\sum_{q i \sigma} f_q  e^{iq x_i} (a_q-a_{-q}^\dagger) n_{i,\sigma}\,,
    \label{eq:S-operator}
\end{equation}
where we use the momentum-space electron density $\rho_{q}=\sum_{i, \sigma} n_{i,\sigma}e^{-i q x_i}$, and the phonon momentum operator $p_{q}=i\sum_{i}( a_{i}^{\dagger}-a_{i})  e^{-i q x_i} /\sqrt{L}$. 

The goal of the NGS-VQE solver is to minimize the total energy in Eq.~\eqref{eq:groundE} in the variational parameter space spanned by $\{f_q\}$, $|\psi_{\rm ph}\rangle$, and $|\psi_{\rm e}\rangle$. Without considering anharmonicity, the phonon state to the right of $U_{\rm NGS}$ should be weakly entangled and can be efficiently captured by variational Gaussian states
\begin{eqnarray}\label{eq:phononwvfuncansatz}
	|\psi_{\rm ph}\rangle = e^{-\frac12 R_0^T \sigma_y \Delta_R} e^{-i\frac14 \sum_\qbf \Rqd \xi_\qbf  \Rq} |0\rangle = \Ub |0\rangle.
\end{eqnarray}
Here, $\Delta_R$, $\xi_\qbf$ are variational parameters and $\Rq = (x_q,p_q)^{\mathrm{T}}$ denotes the bosonic quadrature notation with canonical position $x_q$ and momentum $p_q$, where we adopt the reciprocal representation for the phonon displacement $x_{q}=\sum_{i}  (a_{j}+ a_{j}^\dagger)e^{-iq r_j}/\sqrt{L}$. For convenience, we parameterize the phonon state using the linearization of $\Ub$ named $S_\qbf$, which satisfies $\Ub^\dagger ( x_\qbf, p_\qbf )^T\Ub = S_\qbf ( x_\qbf, p_\qbf )^T$. The NGS-VQE method minimizes the total energy by updating $|\psi_{\rm e}\rangle$ and  $|\psi_{\rm ph}\rangle$ iteratively.

With fixed $U_{\rm NGS}$ and $|\psi_{\rm ph}\rangle$, the electronic problem that the quantum machine has to solve is the ground state of an effective Hamiltonian
\begin{equation}
\begin{aligned}
\label{eq:EH_Hamiltonian}
    \mathcal{H}_{\mathrm{eff}} = &-\tilde{t} \sum_{\langle i,j \rangle, \sigma} \left(c_{i,\sigma}^\dagger c_{j,\sigma}^{} + \mathrm{h.c.}\right) + U \sum_i n_{i,\uparrow}n_{i,\downarrow}\\
    &+ \sum_{i,j} \sum_{\sigma,\sigma'} \tilde{V}_{ij} n_{i,\sigma} n_{j,\sigma'}+ \tilde{E}_{\rm ph}\,, 
\end{aligned}
\end{equation}
where $\tilde{E}_{\rm ph} = \frac14\omega\sum_\qbf\left(\mathrm{Tr} [S_\qbf S_\qbf^\dagger]-2\right)$. The phonon-dressed hopping becomes
\begin{eqnarray}\label{eq:effHopping}
    \tilde{t} = te^{-\sum_\qbf f_\qbf|^2(1-\cos q) e_2^T S_\qbf S_\qbf^\dagger e_2/L},
\end{eqnarray}
and the effective interaction is
\begin{eqnarray}\label{eq:effInteraction}
    \tilde{V}_{ij} = \frac1{L}\sum_{\qbf} \left[2 \omega_0f_\qbf^2-4g_\qbf f_\qbf \right]e^{iq(r_i-r_j)}.
\end{eqnarray}
The VQE solution of the effective Hamiltonian in Eq.~\eqref{eq:EH_Hamiltonian} gives $|\psi_{\rm e}\rangle$ in Eq.~\eqref{eq:wvfuncansatzFullm}. The iterative optimization of $U_{\rm NGS}$ and $|\psi_{\rm ph}\rangle$, for fixed $|\psi_{\rm e}\rangle$, follows the imaginary time evolution in Ref.~\onlinecite{wang2020zero}. It is worth noting that the charge density correlation functions $\langle \rho_{q} \rho_{-q} \rangle \propto \sum_{ij} \sum_{\sigma \sigma'} n_{i,\sigma} n_{j,\sigma'}$ necessary for the imaginary time evolution of $|\psi_{\rm ph}\rangle$ appear in $\mathcal{H}_\mathrm{eff}$ as well. They are therefore already measured with the energy expectation value during VQE and result in no additional computational cost.

\subsection{Quantum circuit and ansatz}

To represent the effective Hamiltonian in Eq.~\eqref{eq:EH_Hamiltonian} on a quantum computer, we rely on the Jordan-Wigner transformation: each electron with given spin orientation is mapped to one qubit. Specifically, it reads
\begin{equation}
    S_{i,\sigma}^{+} = c_{i,\sigma}^\dagger e^{i \pi \sum_{l<i} n_{l,\sigma}},
\end{equation}
\begin{equation}
    S_{i,\sigma}^{-} = c_{i,\sigma}^{} e^{-i \pi \sum_{l<i} n_{l,\sigma}},
\end{equation}
where the phase factors retain the fermionic anti-commutation in the spin operators $S$. Transforming the effective model described by Eq.~\eqref{eq:EH_Hamiltonian} with $L$ sites then yields $2L$ spin operators. Consequently, a quantum ansatz for the electronic ground state contains $2L$ qubits, which represent occupied $| 1 \rangle = c^\dagger |0\rangle$ or unoccupied $| 0 \rangle$ fermionic states. As the effective Hamiltonian preserves occupation number and total spin, we can restrict the electron occupation to half-filling and total spin to zero. Correspondingly, we arrange the qubits representing the two spin orientations separately (see Fig.~\ref{fig:circuit_vqe}\textbf{a}), and require that gates connecting the two spin sectors cannot change their respective occupation.

As outlined in the main text, the employed quantum circuit is based on the Hamiltonian variational ansatz\,\cite{wecker2015progress}. To encode the non-interacting $(U = 0, V_{ij} = 0)$ electronic model, a sequence of Givens rotations $G(\theta)$, parametrized by $\theta$, is applied to adjacent qubits\,\cite{ortiz2001quantum,wecker2015solving,verstraete2009quantum,jiang2018quantum}. In the basis $|00\rangle, |01\rangle, |10\rangle, |11\rangle$, the gate is defined as 
\begin{equation}
G(\theta) = 
    \begin{pmatrix}
        1 & 0& 0 & 0 \\
         0 & \cos\frac{\theta}{2}& -\sin\frac{\theta}{2} & 0 \\
          0 & \sin\frac{\theta}{2}& \cos\frac{\theta}{2} & 0 \\
           0 & 0& 0 & 1 \\
    \end{pmatrix}.
\end{equation}
The ground state of the full effective model is then obtained by an adiabatic evolution with the Hubbard-like Hamiltonian\,\cite{cade2020strategies,stanisic2022observing}. It can be decomposed into kinetic hopping terms,
\begin{equation}
    H(\theta) = e^{-i(c_i^{\dagger}c_{i+1}^{}+c_{i+1}^{\dagger}c_i^{})\theta } = e^{-i(X_i X_{i+1}+Y_i Y_{i+1})\theta/2},
\end{equation}
in the basis $|00\rangle, |01\rangle, |10\rangle, |11\rangle$ represented as 
\begin{equation}
    H(\theta) = \begin{pmatrix}1&0&0&0\\
0&\cos\frac{\theta}{2}&-i\sin\frac{\theta}{2}&0\\
0&-i\sin\frac{\theta}{2}&\cos\frac{\theta}{2}&0\\
0&0&0&1\end{pmatrix},
\end{equation}
and on-site interactions
\begin{equation}
    P(\theta) = e^{-i n_i n_{j}\theta } = e^{-i | 11 \rangle\langle1 1 |_{ij}\theta},
\end{equation}
described by 
\begin{equation}
    P(\theta) = \begin{pmatrix}1&0&0&0\\
0&1&0&0\\
0&0&1&0\\
0&0&0&e^{i\theta}\end{pmatrix}.
\end{equation}
The alternating sequence of phase gates ($P$) and hopping gates ($H$) is then repeated for a number of repetitions $n$, controlling the expressibility of the variational ansatz.

\section*{Data Availability}
The presented data are deposited into the public folder \href{https://figshare.com/projects/A_Hybrid_Quantum-Classical_Method_for_Electron-Phonon_Systems/175065}{Figshare}. Additional numerical data that support the findings of this study are available from the corresponding authors upon reasonable request.

\section*{Code Availability}
The relevant scripts of this study are available from the corresponding authors upon reasonable request.

\section*{Acknowledgments}
We acknowledge technical help from Vivek Dixit and Jiarui Liu and insightful discussions from Kenny Choo and Tao Shi. M.M.D. and T.N. acknowledge support from the European Research Council (ERC) under the European Unions Horizon 2020 research and innovation program (ERC-StG-Neupert-757867-PARATOP). M.M.D. was further funded by a Forschungskredit of the University of Zurich, Grant No. FK-22-085. H.Y. and Y.W. acknowledge support from the National Science Foundation (NSF) awards DMR-2038011 and DMR-2337930. E.D. acknowledges support from the ARO grant number W911NF-20-1-0163 and from the Swiss National Science Foundation under Division II. Simulation results were obtained using the Frontera computing system at the Texas Advanced Computing Center. IBM, the IBM logo, and ibm.com are trademarks of International Business Machines Corp., registered in many jurisdictions worldwide. Other product and service names might be trademarks of IBM or other companies. The current list of IBM trademarks is available at \url{https://www.ibm.com/legal/copytrade}.

\section*{Author contributions} 
Y.W. and E.D. conceived the project. M.M.D., A.M., H.Y., and Y.W. designed, implemented, and tested the code. M.M.D. led the data curation and analysis. M.M.D. and Y.W. wrote the paper with the help from I.T., T.N., and E.D.

\section*{Competing interests} The authors declare no competing interests.

\bibliography{refList}
\clearpage

\setcounter{page}{1}
\setcounter{section}{0}
\setcounter{figure}{0}

\makeatletter
\renewcommand{\@seccntformat}[1]{\csname the#1\endcsname}
\newcommand{\customlabel}[2]{%
   \protected@write \@auxout {}{\string \newlabel {#1}{{#2}{\thepage}{#2}{#1}{}} }%
   \hypertarget{#1}{}
}
\renewcommand\thesection{SUPPLEMENTARY NOTE \arabic{section}:~}
\renewcommand\thesubsection{\Alph{subsection}.~}
\renewcommand\bibname{Supplementary References}
\renewcommand\thefigure{S\arabic{figure}}
\makeatother
\clearpage
\onecolumngrid

\title{Supplementary Information: A Hybrid Quantum-Classical Method for Electron-Phonon Systems}

\maketitle
\newpage
\onecolumngrid

\begin{figure*}[b!]
    \includegraphics[width=\linewidth]{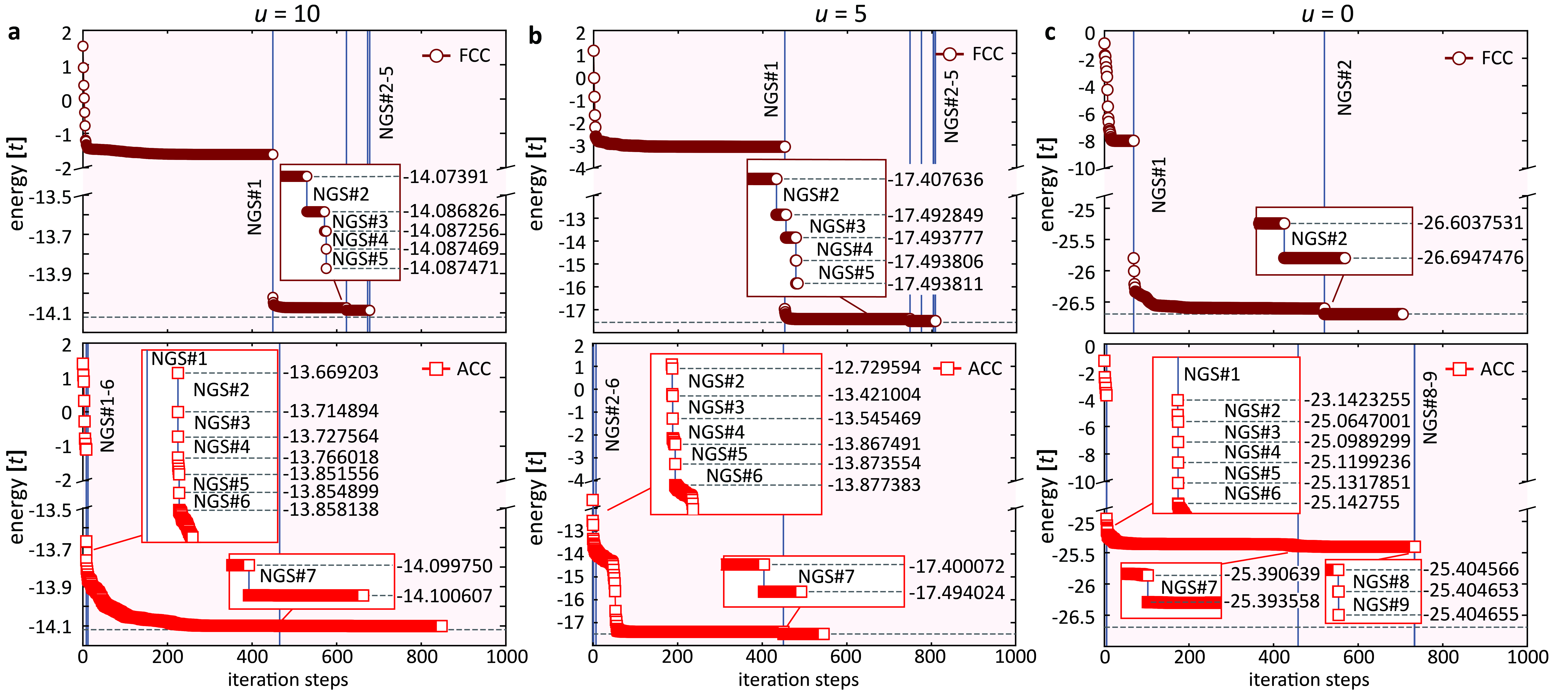}
    \caption{\label{fig:convergence} \textbf{Convergence comparison of VQE approaches.} We compare a VQE algorithm with adaptive energy convergence criterion (ACC, bottom), starting at small accuracy for the first few iterations ($\Delta \epsilon= 10^{-1}$, $10^{-1}$, $10^{-2}$, $10^{-2}$, $10^{-3}$, $10^{-3}$, then $10^{-9}$), with a VQE version of fixed accuracy of $\Delta \epsilon=10^{-9}$ (FCC, top). Inner-loop VQE iterations are displayed in full detail, whereas NGS iterations are compressed. (\textbf{a},\textbf{b}) For large interactions $u = 10, 5$, adaptive convergence speeds up the groundstate search and even improves the overall result when compared to ED. \textbf{c} For vanishing interaction strengths $u$, convergence is slower and worse for the adaptive approach. The phonon solver cannot account for the loss of accuracy in the electronic wavefunction. Consequently, at finite $u$, noisy or less accurate VQE outputs can still lead to precise results, an important criterion for hardware implementations. All panels are for $\omega = 10$, $\lambda = 2$, and $L = 4$ systems and quantum circuit depth $n = 6$.}
\end{figure*}

\section{Adaptive convergence criterion for NGS-VQE iterations}
\label{Supplementary_Note_1}

Hardware implementations are usually limited by the decoherence time of the employed qubit platform. Consequently, limiting the depth of a quantum circuit is crucial, however, this results in less accurate VQE results when solving for the ground state of $\mathcal{H}_{\rm eff}$. Considering the outer-loop NGS-VQE iterations, we propose an adaptive convergence scheme to circumvent this issue: starting from a lower accuracy of the ground-state energy obtained by VQE, we gradually increase the accuracy with the progress of the outer-loop iterations between NGS and VQE. A lower accuracy in the estimated ground state energy corresponds to shallower quantum circuits, which are then gradually deepened to increase the expressibility.

\begin{figure*}[t!]
    \includegraphics[width=\linewidth]{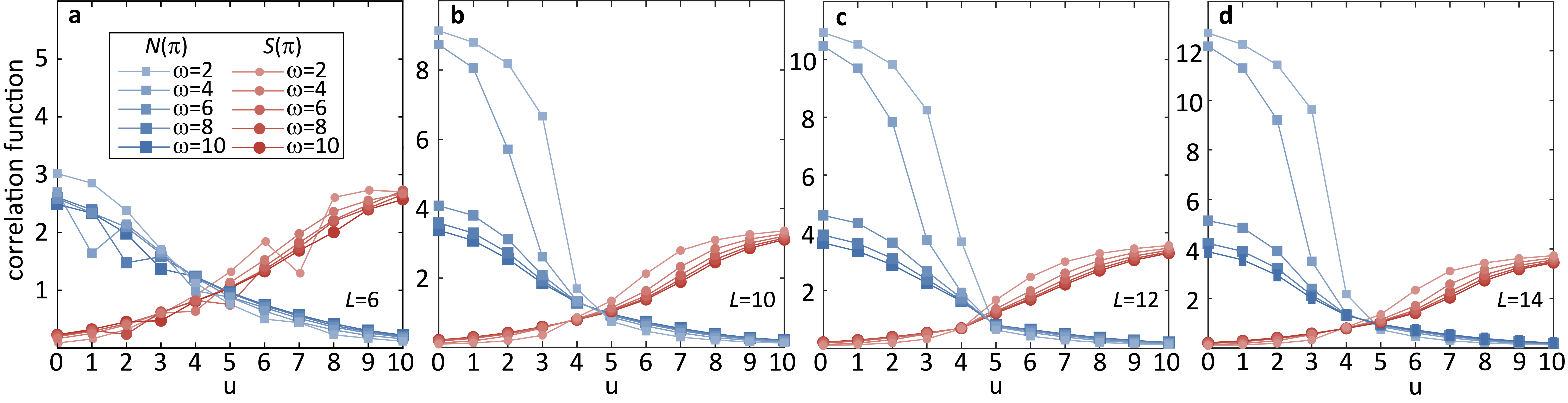}\vspace{-3mm}
    \caption{\label{fig:correlation_functions} \textbf{Correlation functions for different system sizes $L$ and phonon frequencies $\omega$.} \textbf{a} Charge $N(\pi)$ and spin $S(\pi)$ structure factors for different on-site interactions $u$ and phonon frequencies ($\omega=2$, $4$, $6$, $8$, $10$), simulated by the NGS-VQE algorithm for the ground state of a 6-site Hubbard-Holstein model. The coupling strength is fixed to $\lambda=2.5$ for all frequencies. \textbf{b,c,d} Results of $N(\pi)$ and $S(\pi)$ for larger system sizes (\textbf{b} $L=10$, \textbf{c} $L=12$, \textbf{d} $L=14$), calculated by the NGSED algorithm using the same model parameters as panel \textbf{a}.
    }
\end{figure*}

Figure \ref{fig:convergence} shows the comparison between a fixed-accuracy convergence criterion (FCC) and adaptive convergence criterion (ACC) for various model parameters. Here, we start with a VQE accuracy of $\Delta \epsilon =  10^{-1}$, i.e. we terminate the VQE (inner-loop) iterations if the energy between iterations changes by less than $\Delta \epsilon$, i.e., $|E^{\mathrm{VQE}}_{i+1}-E^{\mathrm{VQE}}_{i}|<\Delta \epsilon$. The outer-loop NGS step is triggered using this less accurate electronic state obtained from VQE. Subsequently we increase $\Delta \epsilon$ with each NGS-VQE iteration, until we reach numerical precision at $\Delta \epsilon =  10^{-9}$. The performance of this ACC strategy depends on the phase of the ground state. In an AFM ($u =10$, Fig.~\ref{fig:convergence}\textbf{a}) or intermediate metallic phase ($u =5$, Fig.~\ref{fig:convergence}\textbf{b}), the ultimate electronic ground state is not significantly different from the initial guess (N\'{e}el state). Therefore, the ACC strategy perfectly reproduces the exact solutions, while consuming much less iteration steps to reach the same total-energy accuracy compared to the regular FCC strategy. In contrast, the ACC strategy fails to converge to the correct ground state in the CDW phase ($u =0$, Fig.~\ref{fig:convergence}\textbf{c}). This is because the ground-state electronic configuration in this case has alternative double occupations, which is intrinsically different from the initial electronic state. Therefore, a less accurate VQE solution, in the first few outer-loop iterations, fails to drive the electronic states to a configuration similar to the ground state; this failure further delays the relaxation of phonon states to form alternating distortions and, accordingly, fails to converge to the true ground state within a reasonable number of steps.

In general, the ACC requires more outer-loop NGS-VQE iterations, as the phonon states have to relax to account for the loss in precision. Nevertheless, except for the CDW phase, the ACC strategy requires less VQE iterations in total, with the same accuracy of the electron-phonon ground states. In other words, errors in the electronic part of the hybrid quantum algorithm can be mitigated by the phonon solver, which is conducted on a classical computer and is computationally cheaper.

\section{Impact of the phonon frequency}
\label{Supplementary_Note_2}

The anti-adiabatic limit $u=2\lambda$ boundary controls the energetically favorable single-site electronic configuration and can be used to estimate the transition between CDW and AFM states. However, simulations with larger system sizes have demonstrated that an intermediate Luther-Emery liquid phase emerges near the boundary, where these two insulating instabilities balance\,\cite{clay2005intermediate, fehske2008metallicity, greitemann2015finite}. This intermediate state is driven by the electronic hopping $t$ but is also controlled by the phonon frequencies. Fig.~\ref{fig:correlation_functions}\textbf{a} shows the parameter dependence of $S(\pi)$ and $N(\pi)$ on the phonon frequency. With increasing phonon frequency, the two correlation functions slightly separate from each other, reflecting the underlying intermediate phase. This change is not obvious in the small 6-site system. As a comparison, we further present the simulation results obtained from larger systems (see Fig.~\ref{fig:correlation_functions}). These simulations are conducted using the NGSED instead, due to the limited gates in our NGS-VQE setting. With the increase of system size, the impact of phonon retardation effects at finite frequencies becomes more obvious. The intermediate regime expands with the phonon frequency and suppresses both the charge and spin instabilities in a relatively wide range of parameters.

\begin{figure*}[b!]
    \includegraphics[width=\linewidth]{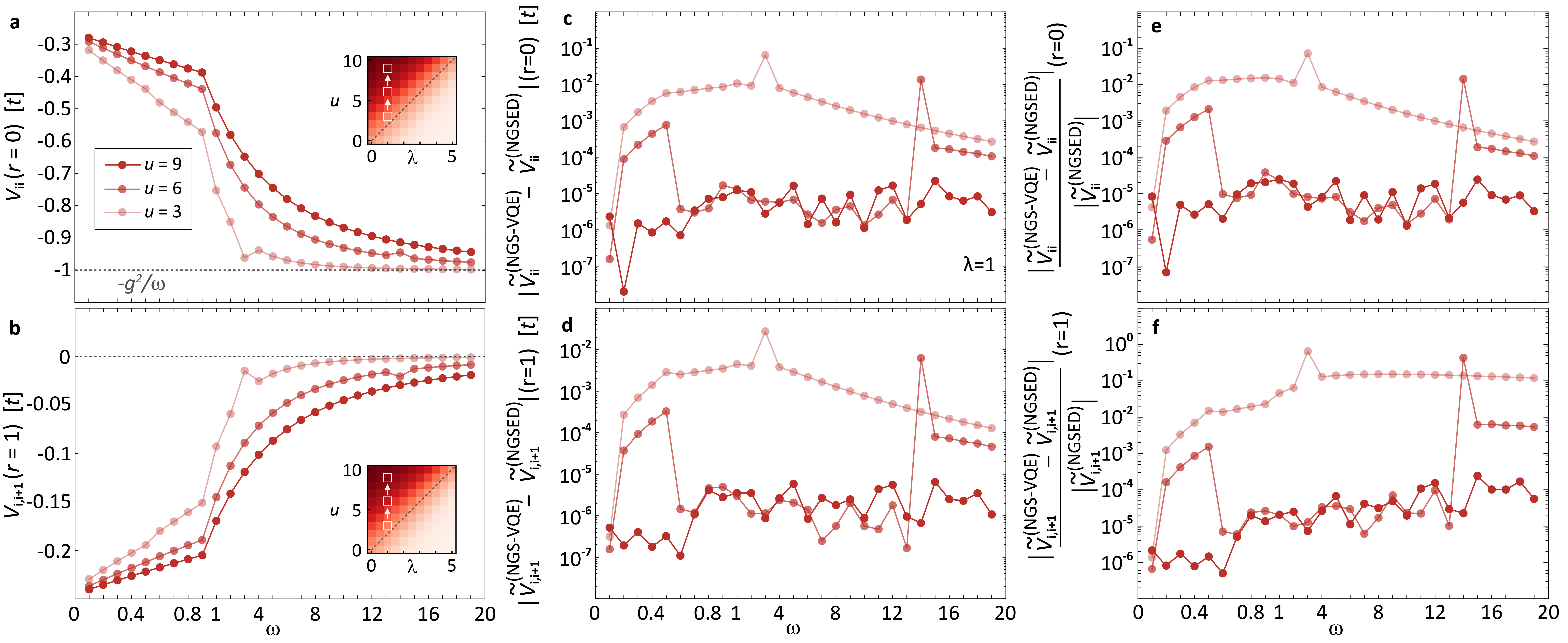}
    \caption{\label{fig:V_vs_g_full} \textbf{Converged extended Hubbard interactions as a function of $\omega$ for $u = 3,6,9$ with fixed $\lambda=1$.} \textbf{a} On-site ($r = 0$) interaction strength $\tilde{V}$ for three $u$ values within the AFM phase ($g' = 0$). While being insensitive to the strength of electronic correlations $u$ for small phonon frequencies, $\tilde{V}$ asymptotically approaches $-\lambda = -1$ in the anti-adiabatic limit ($\omega \rightarrow \infty$). \textbf{b} Nearest neighbor ($r = 1$) interaction strength for three $u$ values within the AFM phase ($g' = 0$). 
    Starting from finite interactions at small phonon frequencies, $\tilde{V}$ quickly reduces in magnitude and vanishes in the anti-adiabatic limit ($\omega \rightarrow \infty$). \textbf{c,e} Absolute and relative error of the on-site ($r = 0$) interaction strength $\tilde{V}$ shown in \textbf{a} as compared to NGSED ($g' = 0$). \textbf{d,f} Absolute and relative error of the nearest-neighbor ($r = 1$) interaction strength $\tilde{V}$ shown in \textbf{b} as compared to NGSED ($g' = 0$).}
\end{figure*}

We further discuss the impact of phonon frequencies on the effective interaction in the Hubbard-Holstein model. As discussed in the Fig.~3 of the main text, $\tilde{V}_{ij}$ approaches $-g^2 \delta_{ij}/\omega$ in the anti-adiabatic limit ($\omega\rightarrow\infty$), while it becomes distributed into a uniform all-to-all interaction in the adiabatic limit ($\omega\rightarrow0$). Here, we show the entire $\omega$ dependence for both the local and nearest-neighbor interactions (see Fig.~\ref{fig:V_vs_g_full}\textbf{a},\textbf{b}). The evolution from adiabatic to anti-adiabatic limits are continuous, without any obvious phase transitions. (Note that the $x$-axis changes its scale at $\omega=1$, which leads to the artificial ``kink'' in the figure.) As mentioned in the main text, the dependence on $u$ becomes less relevant in both limits due to different origins.

In addition, we analyze the relative error of the NGS-VQE simulated $\tilde{V}_{ij}$ against NGSED results. As shown in Fig.~\ref{fig:V_vs_g_full}\textbf{c}-\textbf{f}, largest relative errors for both the local and nearest-neighbor interactions are obtained for a smaller Hubbard interaction $u$ close to the phase transition. This error decreases with increasing $u$. We attribute this error to the sensitivity of the effective interaction $\tilde{V}_{ij}$ to the electronic state. When the system is deeply in the AFM phase, electron density fluctuations are heavily suppressed, although the magnetic configuration stays in a cat state. Since phonons couple to the electron density, the wavefunction is determined to lowest order by the density distribution. Therefore, the phonon-mediated interactions $\tilde{V}_{ij}$ exhibit very small errors for large-$u$ systems. In contrast, the $u=3$ system (with $\lambda=1$) is close to the crossover, where the electron density fluctuation is still strong. Such a fluctuation requires deeper quantum circuits to capture, especially in the shallow AFM side where most gates are used to express the spin configurations. For any strengths of the Hubbard $u$, the relative error for the on-site interactions $\tilde{V}_{ii}$ decreases with the rise of $\omega$, since $\tilde{V}_{ij}$ narrows in space and the absolute value of the on-site interaction increases. In contrast, the relative error for the nearest-neighbor interaction $\tilde{V}_{i,i+1}$ increases with $\omega$, which originates from the same reason (see Fig.~\ref{fig:V_vs_g_full}\textbf{e},\textbf{f}).

\section{Noise simulations}
\label{Supplementary_Note_3}

To assess the practicality of our approach with near-term quantum devices, we study how resilient it is to characteristic hardware noise. To this end, we include both statistical noise ($10^5$ shots per operator expectation value) and realistic device noise in our simulations, specifically, the noise model of IBM's device \textit{ibmq\_kolkata}~\cite{ibmqx}. Scaling the average device errors across several orders of magnitude allows for a detailed understanding of their influence on the solver's accuracy. In particular, our model applies the following noise to all qubits, which are common values for the \textit{ibmq\_kolkata} device at $\eta = 1$:
\begin{equation}
\begin{array}{l}
    T_1 = T_1^{\mathrm{ave}} / \eta 
\ , \quad
T_2 = T_2^{\mathrm{ave}} / \eta 
\ , \quad
e_{\mathrm{1q}} =  \eta e_{\mathrm{1q}}^{\mathrm{ave}}   \\
    e_{\mathrm{2q}} =  \eta e_{\mathrm{2q}}^{\mathrm{ave}}
\ , \quad
e_{\mathrm{ro}} =  \eta e_{\mathrm{ro}}^{\mathrm{ave}}
\end{array}
\end{equation}
where $\eta \in \{ 0.1,0.5,1,5,10 \}$ is a scaling factor and $T_1^{\mathrm{ave}} = 106.1 \, \mathrm{\mu s}$, $T_2^{\mathrm{ave}} = 82.93 \, \mathrm{\mu s}$, $e_{\mathrm{1q}}^{\mathrm{ave}} = 3.78 \times 10^{-4}$, $e_{\mathrm{2q}}^{\mathrm{ave}} = 1.07 \times 10^{-2}$, and $e_{\mathrm{ro}}^{\mathrm{ave}} = 2.27 \times 10^{-2}$ are the device's average relaxation time, dephasing time, one-qubit gate, two-qubit gate, and readout error, respectively.

\begin{figure*}[t!]
    \includegraphics[width=0.7\linewidth]{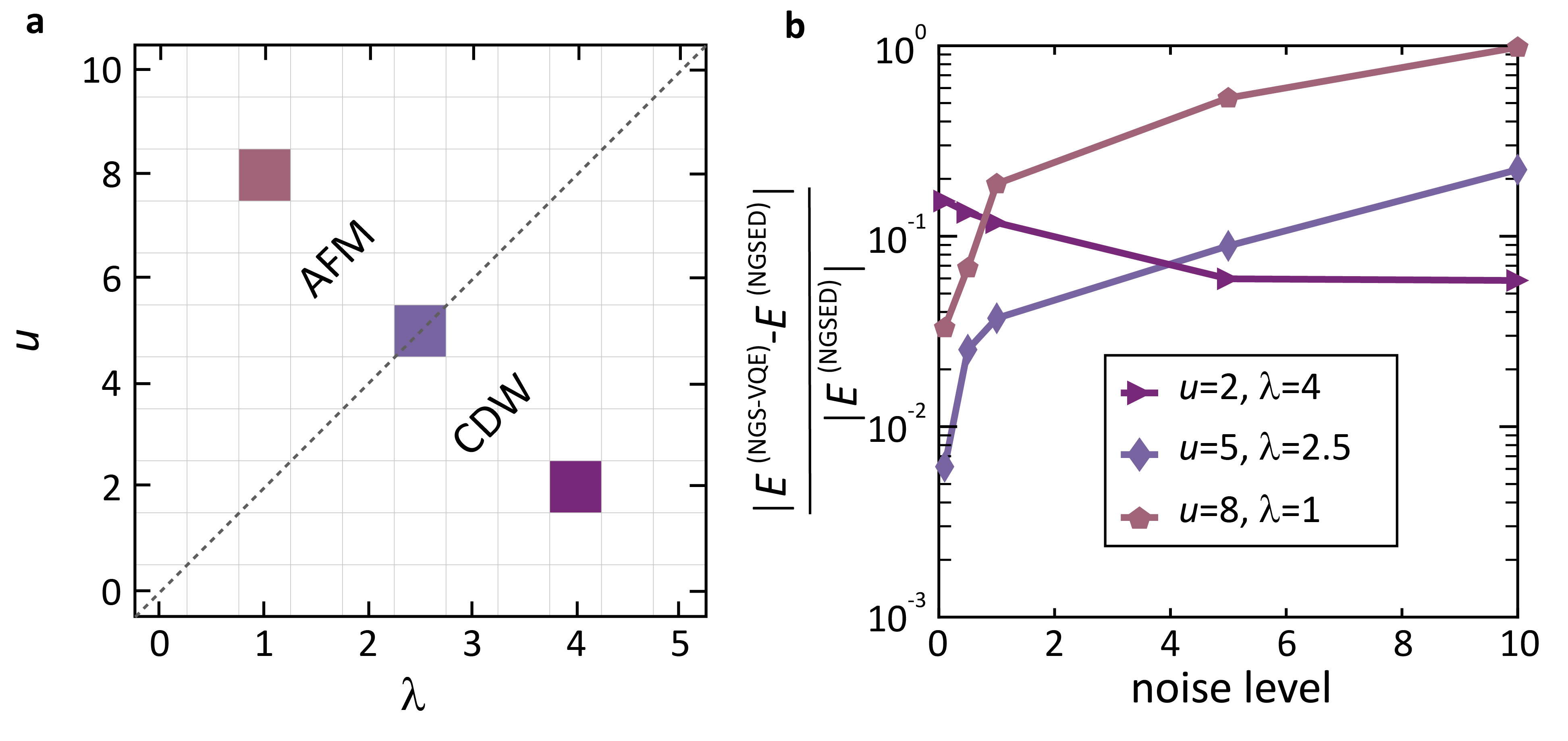}
    \caption{\label{fig:noise_sim} \textbf{Influence of noise on the performance of the hybrid quantum solver.} \textbf{a} Phase diagram of the 1D Hubbard-Holstein model. The highlighted points were used for the noise simulations of panel \textbf{b}. \textbf{b} Relative error of the converged ground state energy for three distinct points in the $u,\lambda$ phase diagram as a function of device noise strength in the VQE solver ($L = 4$, $\omega = 10$). A noise strength of 1 corresponds to a typical configuration of IBM's device \textit{ibmq\_kolkata}. VQE was performed with the circuit depth obtained in Fig.4~\textbf{b} of the main text. While the relative error in the ground state energy increases with increasing noise for the phase transition point and AFM phase, it decreases in the CDW phase.}
\end{figure*}

We observe the expected behavior of an increasing simulation error with increasing levels of device noise in the AFM phase and at the phase transition (see Fig.~\ref{fig:noise_sim}\textbf{b}). Crucially, we obtain $\Delta_\mathrm{rel}(E) = |E_{\mathrm{NGS-VQE}}-E_{\mathrm{NGSED}}|/|E_{\mathrm{NGSED}}|$ at $\eta = 1$ about one order of magnitude larger than in the statevector case (see Fig.~4 of the main text) and dropping off steeply for lower noise levels, as are anticipated with the next generation of quantum devices. For the CDW phase, however, we see a slight decrease in simulation error with increasing hardware noise. This can occur when the optimal solution is such that the noise will naturally relax the system toward this solution, e.g., the qubit ground state $\ket{0\ldots0}$. Additionally, as the CDW phase is dominated by electron-phonon interactions, the NGS solver becomes more important. Fluctuations have proven to be crucial to escape local minima~\cite{wang2020zero}, possibly also explaining the increasing accuracy of the simulation results with noise.  

\end{document}